\documentclass[11pt]{article}
\usepackage{hyperref}
\usepackage{amssymb}
\usepackage{graphicx}
\usepackage{slashed}
\usepackage{amsmath}
\usepackage{tabularx}
\usepackage{breqn}
\usepackage{authblk}
\usepackage{array}
\usepackage{dsfont}
\usepackage{cite}
\usepackage{physics}
\usepackage[section]{placeins}
\usepackage[a4paper, total={6.8in, 10in}]{geometry}
\numberwithin{equation}{section}
\newcommand{\be}{\begin{equation}}
\newcommand{\ee}{\end{equation}}
\newcommand{\bea}{\begin{eqnarray}}
\newcommand{\eea}{\end{eqnarray}}
\newcommand{\ba}{\begin{aligned}}
\newcommand{\ea}{\end{aligned}}
\begin{document}
\title{Maximal acceleration in a Lorentz invariant non-commutative space-time}

\author{E. Harikumar \thanks{eharikumar@uohyd.ac.in}, Suman Kumar Panja
\thanks{sumanpanja19@gmail.com} and Vishnu Rajagopal \thanks{vishnurajagopal.anayath@gmail.com}}
\affil{School of Physics, University of Hyderabad, \\Central University P.O, Hyderabad-500046, Telangana, India}

\maketitle

\begin{abstract}

In this paper, we derive the non-commutative corrections to the maximal acceleration in the Doplicher-Fredenhagen-Roberts (DFR) space-time and show that the effect of the non-commutativity is to decrease the magnitude of the value of the maximal acceleration in the commutative limit. We also obtain an upper bound on the acceleration along the non-commutative coordinates using the positivity condition on the magnitude of the maximal acceleration in the commutative space-time. From the Newtonian limit of the geodesic equation and Einstein's equation for linearised gravity, we derive the explicit form of Newton's potential in DFR space-time. By expressing the non-commutative correction term of the maximal acceleration in terms of Newton's potential and applying the positivity condition, we obtain a lower bound on the radial distance between two particles under the gravitational attraction in DFR space-time. We also derive modified uncertainty relation and commutation relation between coordinates and its conjugate, due to the existence of maximal acceleration.

\end{abstract}

\section{Introduction}

One of the possible approaches to investigate quantum gravity is non-commutative (NC) geometry, which inherently factors in the notion of minimal length scale associated with quantum gravity \cite{connes}. Various models on different NC space-times and their implications have been studied extensively over the past few decades \cite{douglas,chaichian,kappa1,doplicher,carlson,amorim,dfra-review,weight,amorim-1}.

Moyal space-time, a well-explored NC space-time in recent times \cite{douglas}, and its coordinates satisfy the commutation relation,
\be
 [\hat{x}_{\mu},\hat{x}_{\nu}]=i\theta_{\mu\nu},\nonumber
\ee
where $\theta_{\mu\nu}$ is a constant anti-symmetric tensor having the dimension of $(length)^2$. Moyal space-time violates the Lorentz symmetry, and its symmetry algebra has been realised using Hopf algebra \cite{chaichian}. The $\kappa$-deformed space-time is another type of NC space-time, equipped with a Lie algebraic type space-time commutation relation, whose symmetry algebra has also been defined using the Hopf algebra structures \cite{kappa1}.  

Most of the NC space-times violate the Lorentz symmetry. However, it was shown that the Moyal type NC space-times could be generalised such that it becomes Lorentz invariant. This is achieved by assigning a Lorentz transformation for the parameter $\theta_{\mu\nu}$ \cite{doplicher}. This NC parameter has then been promoted to coordinate operators $\hat{\theta}_{\mu\nu}$ \cite{carlson}, and this new space-time is called the DFR space-time. The algebra associated with the DFR space-time has further been extended by including the conjugate momenta $\hat{k}_{\mu\nu}$ corresponding to the new coordinates, $\hat{\theta}_{\mu\nu}$, to form the Doplicher-Fredenhagen-Roberts-Amorim (DFRA) algebra \cite{amorim,dfra-review}. Different aspects of field theories in DFR space-time have been studied in detail in \cite{weight,amorim-1}, and it has been shown that the condition $\hat{\theta}_{0i}=0$ is necessary to ensure the unitarity of field theories in DFR space-time \cite{gomis}. In \cite{guzman} the covariant equations of motion for a particle moving in curved DFR space-time has been derived. The general relativistic field equations and the wave equation for the gravitational field has been derived in DFR space-time \cite{abreu}.  

NC models bring in several features which are not shown by the commutative space-time models and some of these are UV/IR mixing, non-locality, and generalised uncertainty principle (GUP). Another important observation is the existence of an upper limit on the proper acceleration, known as maximal acceleration (MA), associated with the minimal length scale in certain quantum gravity models \cite{kothwalla}. In string theory, it has been shown that there exists a critical acceleration for a given string length \cite{frolov}. The MA has been shown to exist in the covariant loop quantum gravity model \cite{rovelli}. The emergence of MA has also been shown from the causally connected line element in $\kappa$-Minkowski space-time \cite{ganesh}. The MA in commutative space-time has been obtained by incorporating the notion of quantum mechanics into the $8$-dim phase-space of a massive particle \cite{caian-1980}. A similar expression for MA has also been derived from Heisenberg's uncertainty relations \cite{caian-1984}. In \cite{brandt} the MA was obtained by comparing the absolute maximum temperature of thermal radiation \cite{sakharov} with the temperature of the vacuum radiation in an accelerated frame. In a generalised geometric framework, from a classical point of view, by using an approach based on the effective theory of metrics, the existence of MA was shown in \cite{torrome}.

The notion of MA has been used to regularise the UV divergences associated with the quantum fields \cite{nest}. MA has also been used to formulate a new transformation law that makes the $8$-dim line element invariant \cite{scarpetta}. This line-element has further been used to study the dynamics of particles, with MA, in curved backgrounds like Schwarzschild, Reissner-Nordstrom and Kerr space-times \cite{feoli}. The MA dependent $8$-dim line element induces a mass-dependent curvature, and this has been shown to violate the equivalence principle \cite{caian-1990}. It has been shown that
the MA avoids the initial singularity in the standard model of cosmology \cite{gaspaini}. In \cite{kuwata}, using MA, an upper bound on the mass of the Higgs boson was derived. The first-order corrections to the MA in $\kappa$ space-time has been obtained in \cite{kmax1}. The MA has also been used to determine the GUP parameter \cite{luciano}.

We derive the NC corrections to the MA in DFR phase-space in this work. This NC correction term depends on the minimal length scale, acceleration associated with the NC coordinate $\theta_{\mu\nu}$ and the rest mass of the particle. We have shown that the effect of this NC correction term is to decrease the value of the magnitude of the commutative MA. Further, we show that the NC correction term to MA increases when the rest mass of the particle decreases. From the positivity condition on the magnitude of the MA, we have derived an upper bound on the acceleration along the NC coordinate $\theta_{\mu\nu}$. After deriving the Newtonian limit of the geodesic equation in DFR space-time, we obtain the acceleration in DFR space-time in terms of Newtonian potential. Then using the solution to Einstein's field equation in DFR space-time, we calculate the acceleration in terms of Newtonian potential in DFR space-time in the weak field approximation.  

This paper is organised in the following manner. Sec. 2 gives a brief description of the DFR space-time and its algebra. Here we also define the DFRA Poincare algebra and the related Casimir in DFR space-time. In sec. 3, we obtain the line element for the $14$-dim DFR phase-space and derive the MA in DFR space-time from it. In sec. 4, we derive the expression for maximal acceleration in curved DFR space-time. Further, we also derive an expression for the maximal temperature using the MA. In subsec. 4.1, we obtain the geodesic equation for a particle in DFR space-time and obtain its Newtonian limit. In subsec. 4.2, we set up the Einstein equation in DFR space-time and from this, we derive the equations of motion for the linearised gravity in DFR space-time and solve them to get the explicit form of Newton's potential. In the subsec. 4.3, we obtain the explicit form of acceleration along the NC coordinate in terms of Newton's potential. In sec. 5, we give the concluding remarks. Finally, in the appendix. A, we show the possibility of modifying the uncertainty principle between the Minkowski coordinate and its conjugate momenta using the MA expression.

 
\section{DFR space-time}
In this section, we summarise the essential definitions of the DFR space-time coordinates and their underlying algebra. We then discuss the Lorentz generators, DFRA Poincare algebra and Casimir of DFR space-time.  

$\hat{x}_{\mu}$ and $\hat{\theta}_{\mu\nu}$ constitute the DFR space-time coordinate operators where $\hat{\theta}_{\mu\nu}$ is antisymmetric. The conjugate momenta corresponding to DFR space-time coordinates $\hat{x}_{\mu}$ and $\hat{\theta}_{\mu\nu}$ are denoted as $\hat{p}_{\mu}$ and $\hat{k}_{\mu\nu}$, respectively. These coordinates and their conjugate momentas satisfy the following non-vanishing commutation relations,    
\begin{equation}\label{dfra1}
\begin{split}
[\hat{x}_{\mu},\hat{x}_{\nu}]&=i\hat{\theta}_{\mu\nu},~~~~[\hat{x}_{\mu},\hat{p}_{\nu}]=i\eta_{\mu\nu},\\
[\hat{x}_{\mu},\hat{k}_{\nu\lambda}]&=-\frac{i}{2}(\eta_{\mu\nu}\eta_{\lambda\rho}-\eta_{\mu\lambda}\eta_{\nu\rho})\hat{p}^{\rho},\\
[\hat{\theta}_{\mu\nu},\hat{k}_{\rho\lambda}]&=i(\eta_{\mu\rho}\eta_{\nu\lambda}-\eta_{\mu\lambda}\eta_{\nu\rho})
\end{split}
\end{equation}
The above relations form the DFRA algebra \cite{amorim}. The DFR space-time upholds the Lorentz symmetry, and the corresponding Lorentz generators are defined as \cite{amorim}
\begin{equation}\label{dlorentz}
 M_{\mu\nu}=\hat{x}_{\mu}\hat{p}_{\nu}-\hat{x}_{\nu}\hat{p}_{\mu}+\frac{1}{2}\hat{\theta}_{\mu\alpha}\hat{p}^{\alpha}\hat{p}_{\nu}-\frac{1}{2}\hat{\theta}_{\nu\alpha}\hat{p}^{\alpha}\hat{p}_{\mu}-\hat{\theta}_{\mu\lambda}\hat{k}^{~\lambda}_{\nu}+\hat{\theta}_{\nu\lambda}\hat{k}^{~\lambda}_{\mu}.
\end{equation}
The generators of the DRFA Poincare algebra, i.e, $M_{\mu\nu}$, ${\hat p}_\lambda$ and ${\hat k}_{\alpha\beta}$, satisfy the following commutation relations 
\begin{equation}\label{dlorentz1}
\begin{split}
 [M_{\mu\nu},\hat{p}_{\lambda}]&=i(\eta_{\mu\lambda}\hat{p}_{\nu}-\eta_{\nu\lambda}\hat{p}_{\mu}),\\
[M_{\mu\nu},\hat{k}_{\alpha\beta}]&=i(\eta_{\mu\beta}\hat{k}_{\alpha\nu}-\eta_{\mu\alpha}\hat{k}_{\nu\beta}+\eta_{\nu\alpha}\hat{k}_{\beta\mu}-\eta_{\nu\beta}\hat{k}_{\alpha\mu}),\\
[M_{\mu\nu},M_{\lambda\rho}]&=i(\eta_{\mu\rho}M_{\nu\lambda}-\eta_{\nu\rho}M_{\lambda\mu}-\eta_{\mu\lambda}M_{\rho\nu}+\eta_{\nu\lambda}M_{\rho\mu}).
\end{split}
\end{equation}
Quadratic Casimir operator associated with the above defined DFRA Poincare algebra is given as \cite{amorim}
\begin{equation}\label{casimir}
 \hat{P}^2=\hat{p}_{\mu}\hat{p}^{\mu}+\frac{\lambda^2}{2}\hat{k}_{\mu\nu}\hat{k}^{\mu\nu},
\end{equation}
where $\lambda$ is the NC parameter having the dimension of length. 

For the field theories in DFR space-time to satisfy the unitarity conditions, one fixes $\hat{\theta}_{0i}=0$ \cite{gomis} and thus the DFR space-time we consider is a 7-dim space-time. Now onwards we use the definitions $\hat{\theta}_i=\frac{1}{2}\epsilon_{ijk}\hat{\theta}^{jk}$ and $\hat{k}_l=\frac{1}{2}\epsilon_{lmn}\hat{k}^{mn}$. Thus Casimir of the DFR Poincare algebra becomes
\begin{equation}\label{casimir1}
\begin{split}
 \hat{P}^2=&\hat{p}_{\mu}\hat{p}^{\mu}+\lambda^2\hat{k}_{i}^2=-E^2+\textbf{p}^2+\lambda^2\textbf{k}^2.
\end{split}
\end{equation}
We will use Eq.(\ref{casimir1}) to construct the $14$-dim DFR phase-space element in the upcoming section.

The star product \cite{star} between functions $f$ and $g$ in the DFR space-time is defined as
\be
 f(x,\theta)\star g(x,\theta)=e^{\frac{i}{2}\theta^{ij}\partial_{i}\partial_{j}'}f(x,\theta)g(x',\theta)\Big|_{x=x'}, \label{starproduct}
\ee
Note that the above-defined star product will be used to calculate line element, geodesic equation and Einstein's equation in the subsequent sections. 


\section{Maximal acceleration in DFR space-time}

In this section, we first obtain the expression for the line element in the $14$-dimensional phase-space in the DFR framework, from the $7$-dim DFR space-time line element and the DFR dispersion relation. We then derive the expression for the MA in DFR space-time from this $14$-dimensional phase space line-element

The line-element in flat DFR space-time is defined as \cite{abreu}
\be
 ds^2=\Xi_{AB}(\theta)\star dX^A\star dX^B
\label{dfr-line}
\ee
where $X^{A}=(x^{\mu},\theta^i/\lambda)$ are the DFR space-time coordinates, with $A=(\mu,i)=1,2..,7$ (i.e, $\mu=1,2,3,4;~i=5,6,7$), and their derivatives are denoted as $\partial_{B}=(\partial_{\mu}, \lambda^2 \partial_{\theta_{i}})$.

Now using the ansatz for the DFR space-time metric \cite{abreu}
\be
 \Xi_{AB}(\theta)=diag~(\eta_{\mu\nu},e^{-\frac{\theta_{1}^2}{2\lambda^4}},e^{-\frac{\theta_{2}^2}{2\lambda^4}},e^{-\frac{\theta_{3}^2}{2\lambda^4}}),
\label{dfr-mink}
\ee
in Eq.(\ref{dfr-line}) and using the definition of star product, we get the $7$-dim line element in DFR space-time as
\be
 ds^2 = -dt^2+d\textbf{x}^2+\Big(e^{-\frac{\theta_{1}^2}{2\lambda^4}}\Big)\frac{d\theta_{1}^2}{\lambda^2}+\Big(e^{-\frac{\theta_{2}^2}{2\lambda^4}}\Big)\frac{d\theta_{2}^2}{\lambda^2}+\Big(e^{-\frac{\theta_{3}^2}{2\lambda^4}}\Big)\frac{d\theta_{3}^2}{\lambda^2}. 
\label{dfr-line1}
\ee 
In the limit $\lambda\to 0$, since $\lim_{\lambda\to 0}\frac{e^{-\frac{\theta_{i}^2}{2\lambda^4}}}{(2\pi\lambda^2)^{1/2}}=\delta(\theta_i)$, thus the $\theta$ part in the above line element reduces to zero and we recover the $1+3$-dim commutative line element \cite{abreu}.

Now we construct the $14$-dim line element in DFR phase-space, by taking the direct sum of the flat DFR space-time line element given in Eq.(\ref{dfr-line1}) and the energy-momentum relation coming from DFR dispersion relation, as  
\be
 ds^2 = -dt^2+d{\bf{x}}^2+\Big(e^{-\frac{\theta_{1}^2}{2\lambda^4}}\Big)\frac{d\theta_{1}^2}{\lambda^2}+\Big(e^{-\frac{\theta_{2}^2}{2\lambda^4}}\Big)\frac{d\theta_{2}^2}{\lambda^2}+\Big(e^{-\frac{\theta_{3}^2}{2\lambda^4}}\Big)\frac{d\theta_{3}^2}{\lambda^2}+\frac{1}{\mu^4}
\Big(-dE^2+d{\bf{p}}^2+\lambda^2d{\bf{k}}^2\Big), \label{lin-ele}
\ee
where $\mu$ is a parameter having the dimension of mass.

Imposing the condition that the line element between two points are causally connected (i.e, $ds^2\leq 0$) and dividing this resultant inequality with the square of infinitesimal time, i.e, $(dt)^2$, we get 
 \be
 1-v^2-\Big(e^{-\frac{\theta_{1}^2}{2\lambda^4}}\Big)\frac{v_{\theta_1}^2}{\lambda^2}-\Big(e^{-\frac{\theta_{2}^2}{2\lambda^4}}\Big)\frac{v_{\theta_2}^2}{\lambda^2}-\Big(e^{-\frac{\theta_{3}^2}{2\lambda^4}}\Big)\frac{v_{\theta_3}^2}{\lambda^2}+\frac{1}{\mu^4}\bigg[\Big(\frac{dE}{dt}\Big)^2-\Big(\frac{d\bf{p}}{dt}\Big)^2-\lambda^2\Big(\frac{d\bf{k}}{dt}\Big)^2\bigg]\geq 0, \label{timlik1}
\ee 
where we defined $v=\frac{d\bf{x}}{dt}$ as the velocity of the particle in commutative space and $v_{\theta_{i}}=\frac{d\theta_i}{dt}$ as the velocity of the particle in extended $\theta_{i}$ directions. We further divide the above equation by 
$1-v^2-\Big(e^{-\frac{\theta_{1}^2}{2\lambda^4}}\Big)\frac{v_{\theta_1}^2}{\lambda^2}-\Big(e^{-\frac{\theta_{2}^2}{2\lambda^4}}\Big)\frac{v_{\theta_2}^2}{\lambda^2}-\Big(e^{-\frac{\theta_{3}^2}{2\lambda^4}}\Big)\frac{v_{\theta_3}^2}{\lambda^2}$ and re-write $\frac{dE}{dt}$ using the DFR dispersion relation in Eq.(\ref{casimir1}) as $\frac{dE}{dt}=\frac{\bf{p}}{E}\frac{d\bf{p}}{dt}+\lambda^2\frac{\bf{k}}{E}\frac{d\bf{k}}{dt}$. Thus we have
 \be
 1+\frac{\bigg[\Big(\frac{\bf{p}^2}{E^2}-1\Big)\Big(\frac{d\bf{p}}{dt}\Big)^2+\Big(\frac{\lambda^4\bf{k}^2}{E^2}-\lambda^2\Big)\Big(\frac{d\bf{k}}{dt}\Big)^2+2\lambda^2\frac{\bf{pk}}{E^2}\frac{d\bf{p}}{dt}\frac{d\bf{k}}{dt}\bigg]}{\mu^4\bigg(1-v^2-\Big(e^{-\frac{\theta_{1}^2}{2\lambda^4}}\Big)\frac{v_{\theta_1}^2}{\lambda^2}-\Big(e^{-\frac{\theta_{2}^2}{2\lambda^4}}\Big)\frac{v_{\theta_2}^2}{\lambda^2}-\Big(e^{-\frac{\theta_{3}^2}{2\lambda^4}}\Big)\frac{v_{\theta_3}^2}{\lambda^2} \bigg)}\geq 0. \label{timlik3}  
 \ee 
As we know that the DFR space-time preserves Lorentz symmetry, the velocity of particle in DFR space-time is also bounded as $v\leq c$. Therefore the acceleration a particle can achieve is larger when $v<< c$. Thus in order to calculate the MA, we choose an instantaneous rest frame of the particle, i.e., $v=0$ and $v_{\theta_{i}}=0$ \cite{caian-1980}. With this the above Eq.(\ref{timlik3}) becomes
\be
 1+\frac{1}{\mu^4}\Big(-m^2\mathcal{A}_{max}^2-\lambda^2 m^2 a_{\theta}^2\Big)\geq 0,  \label{timlik4}
\ee
where $m\mathcal{A}_{max}=\frac{d\bf{p}}{dt}$ and $ma_{\theta}=\frac{d\bf{k}}{dt}$. Note that here $a_{\theta}$ represents the magnitude of the acceleration (i.e, $a_{\theta}=\sqrt{a_{\theta_1}^2+a_{\theta_2}^2+a_{\theta_3}^2}$) in $\theta$ coordinates, $\lambda^2a_{\theta_i}=\frac{d^2\theta_i}{dt^2}$. By simplifying the above equation, and setting the dimensionfull parameter $\mu$ as $m$ (which is the rest mass of the particle), we obtain the expression for the magnitude of maximal acceleration (in SI units) as
\be
 \mathcal{A}_{max}\leq \frac{mc^3}{\hbar}\Big(1-\frac{\lambda^2 a_{\theta}^2\hbar^2}{2m^2c^6}\Big), \label{timlikf}
\ee 
We observe that the second term is purely due to the non-commutativity, which depends on the square of the minimal length scale $\lambda$. It also depends on the acceleration associated with the extended $\theta$-coordinates, $a_{\theta}$ and is inversly proportional to the rest mass of the particle. Thus the effect of this NC correction term will be more prominent for the lighter particles. It is to be noted that this NC correction term is always a positive quantity. Thus we see that the effect of the non-commutativity is to reduce the magnitude of the MA as compared with the $\mathcal{A}_{max}$ in commutative space-time. A similar decrease in the magnitude of the MA due to non-commutativity was observed in the $\kappa$-deformed space-time \cite{kmax1}.
  
Note that the above equation reduces to $\frac{mc^3}{\hslash}$, the commutative expression for the MA \cite{caian-1980}, in the commutative limit $\lambda\to 0$. As in the commutative space-time, here also the MA diverges, i.e., $\mathcal{A}_{max}\to\infty$, when we take the classical limit $\hslash\to 0$. The non-commutative correction term of $\mathcal{A}_{max}$ is proportional to $\hslash$, and this goes to zero as $\hslash\to 0$.

As mentioned earlier the second term in Eq.(\ref{timlikf}) will always be a positive quantity and the $\mathcal{A}_{max}$, being the magnitude of the proper acceleration is also a positive quantity. This allows us to set an upper bound on the acceleration of the particle in extended $\theta$ direction as
\be 
 \lambda a_{\theta} \leq  \frac{\sqrt{2}mc^3}{\hbar}. \label{bound2}
\ee 
Here we see that as $\hslash\to 0$, $\lambda a_{\theta}\to\infty$, which is similar to the behaviour of $\mathcal{A}_{max}$ in the commutative space-time. Choosing $\lambda$ to be the Planck length, we get a bound on the acceleration of electron moving in $\theta$ direction as $a_{\theta}\leq 2.05\times 10^{64}~s^{-2}$.

The temperature associated with an accelerating particle (of proper acceleration A) can be expressed using the Unruh temperature, i.e, $T_{U}=\frac{\hbar A}{2\pi ck_{B}}$. Using this expression for Unruh temperature, we write down the maximum attainable temperature for a massive particle as $T_{max}=\frac{\hbar \mathcal{A}_{max}}{2\pi ck_{B}}$ \cite{brandt}. Substituting the explicit form of $\mathcal{A}_{max}$, from Eq.(\ref{timlikf}), we get
\be 
 T_{max}=\frac{mc^2}{2\pi k_{B}}\Big(1-\frac{\lambda^2 a_{\theta}^2\hbar^2}{2m^2c^6}\Big). \label{bound5}                                              
\ee
From the above, we find that in the limit $\lambda\to 0$, we get the commutative expression for maximal temperature \cite{brandt}.

Using the experimental observation of the Unruh temperature from the radiation emitted by the positron we get $T=1.80\pm 0.51~PeV$ \cite{morgan}. By setting the error bar of $T$, i.e, $\Delta T=0.51~PeV$ to be equal to the NC correction term of $T_{max}$ in Eq.(\ref{bound5}), we get the bound on $a_{\theta}$ as $a_{\theta}\leq 1.61\times 10^{64}~s^{-2}$.


\section{Maximal acceleration in curved DFR space-time}

In this section, we derive $a_{\theta}$ in the Newtonian limit and obtain the explicit form of the upper cut-off on MA in the Newtonian limit of DFR space-time. For this first, we obtain the Newtonian limit of the geodesic equation in DFR space-time, which gives $a_{\theta}$ in terms of the linearised perturbation $h(x,\theta)$ of the metric of the curved DFR space-time. We then solve Einstein's equation in the weak field approximation and obtain an explicit form of $h(x,\theta)$. Using this, we calculate $a_{\theta}$ and thus $\mathcal{A}_{max}$ in the Newtonian limit. We also show that the positivity of this $\mathcal{A}_{max}$ introduces a lower cut off for the separation of bodies in DFR space-time. 
  
\subsection{Geodesic equation in DFR space-time}  

In this subsection, we construct the geodesic equation of a particle moving in curved DFR space-time and then obtain its Newtonian limit.

We obtain the expression for the geodesic equation in DFR space-time by replacing the usual space-time coordinates with DFR space-time coordinates and the usual product with the star product. Thus we have
\be
\frac{d^2X^{A}}{d\tau^{2}}+\Gamma^{A}_{~BC}*\frac{dX^{B}}{d\tau}*\frac{dX^{C}}{d\tau}=0 \label{gdes1}
\ee
where the Christoffel symbol is defined as 
\be
 \Gamma^{A}_{~BC}=\frac{1}{2}G^{AD}*(\partial_{B}G_{DC}+\partial_{C}G_{BD}-\partial_{D}G_{BC}), \label{christ}
\ee
and $G_{AB}$ is given as \cite{abreu}
\be
 G_{AB}(x,\theta)=diag~(g_{\mu\nu}(x),e^{-\frac{\theta_{1}^2}{2\lambda^4}},e^{-\frac{\theta_{2}^2}{2\lambda^4}},e^{-\frac{\theta_{3}^2}{2\lambda^4}}),
\label{dfr-metric}
\ee
Note $G^{AB}$ is the inverse of $G_{AB}$, satisfying $G_{AB}\star G^{AC}=\delta_B^C$.

In order to derive the Newtonian limit of the above equation we assume that the particle is moving slowly, i.e, $\frac{dx^{i}}{d\tau}<<\frac{dx^{0}}{d\tau},~\frac{d\theta^{i}}{d\tau}<<\frac{dx^{0}}{d\tau}$, and the gravitational field is static i.e, $\frac{\partial {G}_{AB}}{\partial t}=0$. Further we also linearise the DFR metric as ${G}_{AB}(x,\theta)={\Xi}_{AB}(\theta)+{h}_{AB}(x,\theta)$ where we assume $|{h}_{AB}|<<1$.

Using the defintion of the star product and the above approximations in Eq.(\ref{gdes1}), we get the following equations,
\be
 \frac{d^{2}t}{d\tau^{2}}=0,\label{gdesf0}
\ee
\be
 \frac{d^{2}x_{i}}{d\tau^{2}}=\frac{1}{2}\Big(\frac{dt}{d\tau}\Big)^2\partial_{i}h_{00}(x,\theta), \label{gdesf1}
\ee
\be
 \frac{d^2\theta_{i}}{d\tau^{2}}=\frac{\lambda^2}{2}\Big(\frac{dt}{d\tau}\Big)^2\partial_{\theta_i}h_{00}(x,\theta). \label{gdes4} 
\ee
From Eq.(\ref{gdesf0}), we find that $\frac{dt}{d\tau}$ is a constant. Thus dividing Eq.(\ref{gdesf1}) and Eq.(\ref{gdes4}) throughout by $\Big(\frac{dt}{d\tau}\Big)^2$, we get the Newton's force equation in DFR space-time as
\be
 \ddot{x}_{i}=\frac{1}{2}\partial_{i}h_{00}(x,\theta), \label{gdesf2}
\ee
\be
 \ddot{\theta}_{i}=\frac{\lambda^2}{2}\partial_{\theta_i}h_{00}(x,\theta), \label{gdes5} 
\ee
where $``\cdot"$ stands for differential with respect to $t$.

Note that here we have $\lambda^2a_{\theta_i}=\ddot{\theta}_i$. We can combine Eq.(\ref{gdesf1}) and Eq.(\ref{gdesf2}) to obtain a covariant form for the Newtonian limit of the geodesic equation of a particle moving in DFR space-time as $\ddot{X}_{A}=\frac{1}{2}\partial_{A}h_{00}(x,\theta)$.

In order to find the explicit form of the Newton's equation from Eq.(\ref{gdesf2}) and Eq.(\ref{gdes5}), we need to know the explicit form of the Newtonian potential $h_{00}(x,\theta)$ in DFR space-time. In the next subsection we will evaluate the explicit form of $h_{00}(x,\theta)$ and then we will derive $a_{\theta_i}$ in terms of $h_{00}(x,\theta)$. 

\subsection{Linearised gravity in DFR space-time}

In this subsection, we first set up the Einsteins field equation in DFR space-time, solve the equations of motion for the linearised metric in DFR space-time and obtain Newton's potential. The Newton's potential is obtained with the additional NC dimensions $1,2,\&3$, separately.

The Einstein's field equation in DFR space-time is defined as \cite{abreu}
\be
 R_{AB}-\frac{1}{2}R\star G_{AB}=16G^{(d_{\theta})}\pi T_{AB} \label{gren0}
\ee  
where $G_{AB}(x,\theta)$ is given in Eq.(\ref{dfr-metric}) and Ricci tensor and scalar are defined as
\be
 R_{AB}=\partial_C \Gamma^{C}_{~AB}-\partial_{A}\Gamma^{C}_{~BC}+\Gamma^{C}_{~CD}\star\Gamma^{D}_{~AB}-\Gamma^{C}_{~AD}\star\Gamma^{D}_{~CB}\label{riccit}
\ee 
and 
\be
 R=R_{AB}\star G^{AB}, \label{riccis}
\ee
respectively. In Eq.(\ref{gren0}), $T_{AB}$ is the energy momentum tensor associated with the matter content in the DFR space-time. Note that $(G^{(d_{\theta})})$ in Eq.(\ref{gren0}) represents the Newton's gravitational constant in $1+3+d_{\theta}$ dimensional DFR space-time, which is defined as $G^{(d_{\theta})}=G\lambda^{d_{\theta}}$. Here $d_{\theta}$ represents the number of $\theta$ dimension in the DFR space-time which can take the values $d_{\theta}=1,2,3$. 

Now we set up the equations of motion, using Eq.(\ref{gren0}), for the linearised metric $h_{AB}$. Choosing harmonic gauge where $\partial_A h=2\partial^C h_{AC}$ and also assuming that the metric is traceless, i.e, $h=h_A^{~A}=0$, Eq.(\ref{gren0}) becomes
\be
 \Big(\square+\lambda^2\square_{\theta}\Big)h_{AB}(x,\theta)=16G^{(d_{\theta})}\pi T_{AB}. \label{gren1}
\ee
The above equation represents the wave equation for the linearised gravitational field in DFR space-time. Unlike the Moyal space-time, here the field propagates in the extra $\theta$ dimensions also. We get the commutative equations of motion in the limit $\lambda\to 0$. 
 
We now solve the above inhomogenous differential equation using the Green's function method. For this we write down the general expression for this inhomogenous solution as  
 \be
 h_{AB}(x,\theta)=\int d^4x'd^{(d_{\theta})}\theta'~W^{(d_{\theta})}(\theta')16G^{(d_{\theta})}\pi T_{AB}(x',\theta')\mathcal{G}^{(d_{\theta})}(X-X'). \label{gren2}
 \ee
The weight function is defined as $W^{(d_{\theta})}(\theta)=\Big(\frac{1}{2\pi \lambda^4}\Big)^{\frac{d_{\theta}}{2}}e^{-\frac{\theta^2}{2\lambda^4}}$ (see \cite{weight} for more details on weight function) and $\mathcal{G}^{(d_{\theta})}(X-X')$ is the Green's function corresponding to Eq.(\ref{gren1}) in $1+3+d_{\theta}$ dimension.
 
In general, for an odd spatial dimension (i.e., for $1+D$ dimensional space-time with odd value of D) we have the expression for the Green's function as \cite{hassani}
 \be
 \mathcal{G}(r,t)=\frac{1}{4\pi}\Big(-\frac{1}{2\pi r}\frac{\partial}{\partial r}\Big)^{\frac{(D-3)}{2}} \frac{\delta(t-r)}{r}. \label{grenodd}
 \ee
Similarly for an even spatial dimension (i.e., for even value of D), we have \cite{hassani}
 \be
 \mathcal{G}(r,t)=\frac{1}{2\pi}\Big(-\frac{1}{2\pi r}\frac{\partial}{\partial r}\Big)^{\frac{(D-2)}{2}} \frac{\Theta(t-r)}{\sqrt{t^2-r^2}}. \label{greneven}
 \ee
Now let us split the spatial dimension of the DFR space-time into usual Minkowskian spatial part and the extra $\theta$ dimensionals as  $D=3+d_{\theta}$. Thus from above, we find the Green's functions in DFR space-time as
\be
 \mathcal{G}(r,t)=\frac{1}{4\pi}\Big(-\frac{1}{2\pi r}\frac{\partial}{\partial r}\Big)^{\frac{(d_{\theta})}{2}} \frac{\delta(t-r)}{r}, ~~~\textnormal{for}~~ d_{\theta}=2 \label{grenodd1}
\ee 
 and 
\be
 \mathcal{G}(r,t)=\frac{1}{2\pi}\Big(-\frac{1}{2\pi r}\frac{\partial}{\partial r}\Big)^{\frac{(d_{\theta}+1)}{2}} \frac{\Theta(t-r)}{\sqrt{t^2-r^2}}, ~~~\textnormal{for}~~ d_{\theta}=1,3  \label{greneven1}
 \ee
We use the explicit form of the Green's functions in different dimensions and evaluate the integral in Eq.(\ref{gren2}), deriving the explicit form of Newton's potential in $d_{\theta}=1,2,3$ dimensions, separately.  
 
\underline{\bf{For $d_{\theta} = 1$}}: Using Eq.(\ref{greneven1}), we find the expression for Green's function in $1+3+1$ dimension as
\be
 \mathcal{G}(X-X')= -\frac{1}{4\pi^2 \Delta X}\frac{\partial}{\partial (\Delta X)}\bigg( \frac{\Theta(\Delta t-\Delta X)}{\sqrt{\Delta t^2-\Delta X^2}}\bigg) \label{gren1a}
\ee
where $\Delta t=t-t'$ and $\Delta X=|\textbf{X}-\textbf{X}'|$. Next we re-express the RHS of the above equation in terms of derivative with respect to $\Delta t$. For this we notice
\bea
 \frac{\partial}{\partial (\Delta X)}\bigg( \frac{\Theta(\Delta t-\Delta X)}{\sqrt{\Delta t^2-\Delta X^2}}\bigg) & = & -\frac{\partial}{\partial (\Delta t)}\bigg( \frac{\Theta(\Delta t-\Delta X)}{\sqrt{\Delta t^2-\Delta X^2}}\bigg)-\bigg( \frac{\Theta(\Delta t-\Delta X)}{(\Delta t^2-\Delta X^2)^{\frac{3}{2}}}\bigg)(\Delta t-\Delta X) \label{gren1b}
\eea
Substituting Eq.(\ref{gren1b}) in Eq.(\ref{gren1a}), we find the explict form of the Green's function as
\be
 \mathcal{G}(X-X')=\frac{1}{4\pi^2 \Delta X}\bigg[\frac{\partial}{\partial (\Delta t)}\bigg( \frac{\Theta(\Delta t-\Delta X)}{\sqrt{\Delta t^2-\Delta X^2}}\bigg)+\bigg( \frac{\Theta(\Delta t-\Delta X)}{(\Delta t^2-\Delta X^2)^{\frac{3}{2}}}\bigg)(\Delta t-\Delta X)\bigg]. \label{gren1c}
\ee
From now onwards we consider a static matter source, i.e,  $T_{00}(x',\theta')=T_{00}(\textbf{X}')$. Thus for $d_{\theta}=1$, the linearised metric in Eq.(\ref{gren2}) becomes
\be
 h_{00}(X)=\int d^3\textbf{x}'d\theta' W^{(1)}(\theta')16G^{(1)}\pi T_{00}(\textbf{x}',\theta')\int dt' \mathcal{G}(X-X') \label{gren1d}
\ee
Now we use Eq.(\ref{gren1c}) and evaluate the time integral of Green's function appearing in the above as \cite{yuchen}
\bea
 \int dt' \mathcal{G}(X-X') = \int^{\infty}_{0} dt'\frac{1}{4\pi^2 \Delta X}\bigg[\frac{\partial}{\partial (\Delta t)}\bigg( \frac{\Theta(\Delta t-\Delta X)}{\sqrt{\Delta t^2-\Delta X^2}}\bigg)+\bigg( \frac{\Theta(\Delta t-\Delta X)}{(\Delta t^2-\Delta X^2)^{\frac{3}{2}}}\bigg)(\Delta t-\Delta X)\bigg] 
\eea 
After changing the integration variable from $t'$ to $\Delta t$, and the integration limits change from $t' \rightarrow 0$ to $\Delta t \rightarrow t$ and $t' \rightarrow \infty$ to $\Delta t \rightarrow -\infty$, we find
\bea
  \int dt' \mathcal{G}(X-X') &=& \frac{1}{4\pi^2 \Delta X}\bigg[\frac{\Theta(\Delta t-\Delta X)}{\sqrt{\Delta t^2-\Delta X^2}}\bigg]^{-\infty}_{t} +\frac{1}{4\pi^2 \Delta X}\int^{-\infty}_{t}d(\Delta t)\frac{\Delta t -\Delta X}{(\Delta t^2 -\Delta X^2)^{\frac{3}{2}}} \\
&=& -\frac{1}{4\pi^2 \Delta X}\frac{\Theta( t-\Delta X)}{\sqrt{t^2-\Delta X^2}} +\frac{1}{4\pi^2 \Delta X}\int^{-\infty}_{t}d(\Delta t)\frac{\Delta t -\Delta X}{(\Delta t^2 -\Delta X^2)^{\frac{3}{2}}}.
\label{gren1e}
 \eea
Now we evaluate the above expression in the limit $t\to\Delta X$. Using the L'Hospital's rule, we find that the first term of Eq.(\ref{gren1e}) vanishes and evaluating the integral in the second term on RHS, we get
\be
  \int dt' \mathcal{G}(X-X')=\frac{1}{\pi^2}\frac{1}{|\textbf{X}-\textbf{X}'|^2}. \label{gren1f}
\ee
From now onwards we use the approximation $|\textbf{X}-\textbf{X}'|^{-1}\approx\frac{1}{r_{\theta}}$ to the lowest order in $\frac{1}{|\textbf{X}|}$ (where $r_{\theta}=|\textbf{X}|$. Note that $r_{\theta}$ is the radial distance in $1+3+d_{\theta}$ dimensional DFR space-time and this depends on $\theta$. In the limit, i.e, $\lim\lambda\to 0$, the radial distance in DFR space-time reduces to the radial distance in commutative space-time, i.e, $r_{\theta}\to r$). Using this and Eq.(\ref{gren1f}) in Eq.(\ref{gren1d}), we get the Newton's potential in $5$-dim DFR space-time as
 \be
 h_{00}(r_{\theta})=\frac{4G^{(1)}M}{\pi}\frac{1}{r_{\theta}^2} \label{gren1i}
 \ee
where $M$ corresponds to the mass of the matter present in this $5$-dim DFR space-time and it is defined as $M=\int d^3\textbf{x}'d\theta'~W^{(1)}(\theta')T_{00}(\textbf{x}',\theta')$. 

\underline{\bf{For $d_{\theta} = 2$}}: From Eq.(\ref{grenodd1}), we get the explicit form of the Green's function in $1+3+2$ dimensional DFR space-time as
\be
 \mathcal{G}(X-X')=\frac{1}{4\pi^2}\frac{\delta(t-t'-|\textbf{X}-\textbf{X}'|)}{|\textbf{X}-\textbf{X}'|^3}. \label{gren0j}
\ee
For $d_{\theta}=2$, the expression for the Newton's potential in Eq.(\ref{gren2}) becomes
\be
 h_{00}(X)=\int d^3\textbf{x}'d^2\theta' W^{(2)}(\theta')16G^{(2)}\pi T_{00}(\textbf{x}',\theta')\int dt' \mathcal{G}(X-X'). \label{gren0j1}
\ee
Substituting Eq.(\ref{gren0j}) in Eq.(\ref{gren0j1}), we get the explicit form of the Newton's potential in $6$-dim DFR space-time as
\be
 h_{00}(r_{\theta})=\frac{4G^{(2)}M}{\pi}\frac{1}{r_{\theta}^3}. \label{gren0j2}
\ee
The mass M in the $6$-dim DFR space-time is defined as $M=\int d^3\textbf{x}'d^2\theta'~W^{(2)}(\theta')T_{00}(\textbf{x}',\theta')$. 

\underline{\bf{For $d_{\theta} = 3$}}: In $1+3+3$ dimensional DFR space-time, we get the expression for the Green's function from Eq.(\ref{greneven1}), as
\be
 \mathcal{G}(X-X')=\frac{1}{8\pi^3}\frac{1}{\Delta X}\frac{\partial}{\partial (\Delta X)}\bigg(\frac{1}{\Delta X}\frac{\partial}{\partial (\Delta X)}\Big(\frac{\Theta(\Delta t-\Delta X)}{\sqrt{\Delta t^2-\Delta X^2}}\Big)\bigg). \label{gren0j3}
\ee 
We now re-write the derivatives in terms of $\Delta t$ (as we did in the case of $d_{\theta}=1$) and we get
\begin{equation}\label{gren0j4}
\begin{split}
 \mathcal{G}(X-X')=&\frac{1}{8\pi^3}\frac{1}{\Delta X^3}\bigg[\frac{\partial}{\partial (\Delta t)}\Big(\frac{\theta(\Delta t-\Delta X)}{\sqrt{\Delta t^2-\Delta X^2}}\Big)+\frac{\theta(\Delta t-\Delta X)}{(\Delta t^2-\Delta X^2)^{3/2}}(\Delta t-\Delta X)\bigg]\\
&-\frac{1}{8\pi^3}\frac{1}{\Delta X^2}\frac{\partial}{\partial (\Delta X)}\bigg[\frac{\partial}{\partial (\Delta t)}\Big(\frac{\theta(\Delta t-\Delta X)}{\sqrt{\Delta t^2-\Delta X^2}}\Big)+\frac{\theta(\Delta t-\Delta X)}{(\Delta t^2-\Delta X^2)^{3/2}}(\Delta t-\Delta X)\bigg]
\end{split}
\end{equation} 
For $d_{\theta}=3$, Eq.(\ref{gren2}) takes the form
\be
h_{00}(X)=\int d^3\textbf{x}'d^3\theta' W^{(3)}(\theta')16G^{(3)}\pi T_{00}(\textbf{x}',\theta')\int dt' \mathcal{G}(X-X'). \label{gren0j5}
\ee
We now follow the same steps as we did in obtaining $h_{00}(X)$ in $d_{\theta}=1$ dimension, and find the Newton's potential in $7$-dim DFR space-time as
\be
 h_{00}(r_{\theta})=\frac{4G^{(3)}M}{\pi^2}\frac{1}{r_{\theta}^4}, \label{gren0j6}
\ee
where $M=\int d^3\textbf{x}'d^3\theta'~W^{(3)}(\theta')T_{00}(\textbf{x}',\theta')$. 

From Eq.(\ref{gren1i}), Eq.(\ref{gren0j2}) and Eq.(\ref{gren0j6}) we see that $h_{00}(\textbf{x},\theta)$ depends on the NC dimensions. Thus, in general, the Newtonian potential, $h_{00}(\textbf{x},\theta)$, in DFR space-time can be written in the SI units as \footnote{Note that here we have absorbed the factors of $\pi$ into the definition of $M$}
\be
 h_{00}(r_{\theta})=\frac{4G^{(d_{\theta})}M}{r_{\theta}^{1+d_{\theta}}} ~~~\textnormal{for}~d_{\theta}=1,2,3. \label{linmetf}
\ee
Note that similar $\frac{1}{r^{1+d_{\theta}}}$ dependence in the Newtonian potential has been observed in the study of the gravity in extra dimensions \cite{extra}.
 
Next substituting Eq.(\ref{linmetf}) in $\ddot{X}_{A}=\frac{1}{2}\partial_{A}h_{00}(x,\theta)$, we get the Newton's force experienced by a particle of mass $\mathcal{M}$ in DFR space-time as
\be
 F_{A}=-\frac{2(1+d_{\theta})G^{(d_{\theta})}M\mathcal{M}X_A}{r_{\theta}^{3+d_{\theta}}}, \label{Newton-fin}
\ee
where $F_{A}=\mathcal{M}\ddot{X}_A$. We find that the Newton's force equation in DFR space-time depends on the radial distance. Similar radial dependency has also been observed in the study of the Newton's force equation in NC space-time \cite{hari}. 

Using $h_{00}(r_{\theta})$, i.e, Eq.(\ref{linmetf}) in Eq.(\ref{gdes5}), we calculate $\lambda^{-2}\ddot{\theta_i}=a_{\theta_i}$ and find that
\be
 a_{\theta_i}=\frac{-2G^{(d_{\theta})}M(1+d_{\theta})\theta_i\lambda^{-2}}{r_{\theta}^{3+d_{\theta}}}.
\ee

\subsection{MA in terms of Newton's potential}

Now we substitute this $a_{\theta}$ in the expression for MA, in, Eq.(\ref{timlikf}), and get the maximal acceleration as
\be
 \mathcal{A}_{max}\leq\frac{mc^3}{\hslash}\bigg(1-\frac{2(1+d_{\theta})^2G^2M^2\lambda^{2(d_{\theta}-1)}\hslash^2\theta^2}{m^2c^6r_{\theta}^{6+2d_{\theta}}}\bigg),
\ee
where $r_{\theta}$ is the radial distance between the gravitating body of mass $M$ and the particle of mass $m$ in DFR space-time. Using the postivity condition of the magnitude of the MA we get a lower bound on this radial distance in DFR space-time as
\be
 r_{\theta}^{6+2d_{\theta}}\geq\frac{2G^2M^2\hslash^2(1+d_{\theta})^2\theta^2\lambda^{2(d_{\theta}-1)}}{m^2c^6}.
\ee 
Here we notice that this lower bound on the radial distance vanishes in both commutative ($\lambda\to 0$) and classical ($\hslash\to 0$) limits. 

The mass term $M$ in the above expressions represent the mass of the total matter in the DFR space-time. When we take the commutative limit, this mass term should also naturally reduce to the mass content associated with the commutative Minkowski space-time. In order to obtain this smooth commutative limit, we re-write the definition of the mass term in terms of matter density. Thus we write the mass term $M$ as  
\be
 M=\int d^3\textbf{x}d^{(d_{\theta})}\theta~W^{(d_{\theta})}(\theta)\rho^{(d_{\theta})}(\textbf{x},\theta) \label{gren1j}
\ee
 where $\rho^{(d_{\theta})}(\textbf{x},\theta)$ is matter density of the $1+3+d_{\theta}$ dimensional DFR space-time. Here we consider an ansatz for this matter density as $\rho^{(d_{\theta})}(\textbf{x},\theta)=\bar{\rho}(\textbf{x})\tilde{\rho}^{(d_{\theta})}(\theta)$ and we set $m=\int d^3\textbf{x}~\bar{\rho}(\textbf{x})$ as the mass content in the $1+3$ dimensional commutative Minkowski space-time. Using this in Eq.(\ref{gren1k}) we find 
\be
 M = m\int d^{(d_{\theta})}\theta~W^{(d_{\theta})}(\theta) \tilde{\rho}^{(d_{\theta})}(\theta). \label{gren1k}
\ee
From the above we observe that in the commutative limit we get $\lim_{\lambda\to 0}\int d^{(d_{\theta})}\theta~W(\theta) \tilde{\rho}^{(d_{\theta})}(\theta)=\int d^{(d_{\theta})}\theta \delta^{(d_{\theta})}(\theta) \tilde{\rho}^{(d_{\theta})}(\theta)=1$ with $\tilde{\rho}^{(d_{\theta})}(0)=1$. Now we obtain the mass term defined in Eq.(\ref{gren1k}) for different possible choices of $\tilde{\rho}^{(d_{\theta})}(\theta)$ such that $\tilde{\rho}^{(d_{\theta})}(0)=1$ and $\tilde{\rho}^{(d_{\theta})}(\theta)$ is Lorentz invariant. We then substitute this in Eq.(\ref{linmetf}) and evaluate the Newtonian potential. Using this we calculate $\ddot{x}_i$ and $\lambda^{-2}\ddot{\theta}_i=a_{\theta_i}$ explicitly from Eq.(\ref{gdesf2}) and Eq.(\ref{gdes5}). These results are summarised in the following tables.

\begin{table}[h!]
\centering
\renewcommand{\arraystretch}{2} 
\begin{tabular}{|c |c |c |c |r |r |r |r |}
\hline
$\tilde{\rho}^{(d_{\theta})}(\theta)$ &  $h_{00}(r_{\theta})$ & $\ddot{x}_{i}$ & $a_{\theta_i}$ \\[2ex]
\hline
$\sum_{n=0}^{\infty}a_{2n}\theta^{2n}$ & $\frac{4G^{(d_{\theta})}m\sum_{n=0}^{\infty}a_{2n}\lambda^{4n}(1+d_{\theta})}{r_{\theta}^{1+d_{\theta}}}$  & $-\frac{2G^{(d_{\theta})}m\sum_{n=0}^{\infty}a_{2n}\lambda^{4n}(1+d_{\theta})x_i}{r_{\theta}^{3+d_{\theta}}}$  & $-\frac{2G^{(d_{\theta})}m\lambda^{-2}\sum_{n=0}^{\infty}a_{2n}\lambda^{4n}(1+d_{\theta})\theta_i}{r_{\theta}^{3+d_{\theta}}}$ \\[2.5ex]
\hline
$e^{-\frac{b\theta^2}{2}}$ & $\frac{4G^{(d_{\theta})}m(1+d_{\theta})}{\sqrt{1+b\lambda^4}}\frac{1}{r_{\theta}^{1+d_{\theta}}}$ & $-\frac{2G^{(d_{\theta})}m(1+d_{\theta})}{\sqrt{1+b\lambda^4}}\frac{x_i}{r_{\theta}^{3+d_{\theta}}}$ & $-\frac{2G^{(d_{\theta})}m\lambda^{-2}(1+d_{\theta})}{\sqrt{1+b\lambda^4}}\frac{\theta_i}{r_{\theta}^{3+d_{\theta}}}$   \\[2.5ex]
\hline
\end{tabular}
\caption{For $d_{\theta}=1,2,3$}
\label{table-1}
\end{table}

From the table. 1 we observe that in the commutative limit (i.e, $\lambda\to 0$), Newton's potential reduces to the usual Newton's potential in $1+3$-dim as, $h_{00}(r_{\theta})\to h_{00}(r)=\frac{4Gm}{r}$, and the acceleration along the Minkowski space coordinates ($\ddot{x}_i$) reduce as $\ddot{x}_i=-\frac{2Gmx_i}{r^3}$, as expected. Note that the acceleration along the extra $\theta$ direction ($a_{\theta}$) vanishes in the commutative limit, as expected.


\section{Conclusion}

In this paper, we have obtained the non-commutative corrections to the maximal acceleration for a massive particle in the DFR space-time. This is derived from the causally connected $14$-dim DFR phase space line element, which is constructed using the $7$-dim DFR space-time metric and the DFR dispersion relation. The NC correction term depends on the square of the minimal length scale and acceleration along the $\theta$-coordinates. Due to the inverse dependence on the rest mass of the particle, the non-commutative correction to the MA is more prominent for the lighter particles. We find that the effect of the non-commutativity is to reduce the magnitude of the MA compared to that in the commutative space-time. This trend was observed in the $\kappa$-Minkowski space-time \cite{kmax1} also. By using this MA and the Unruh temperature, we have derived the maximum attainable temperature for a massive particle in DFR space-time. From both the experimental observations of the Unruh radiation and the positivity condition on the magnitude of the MA, we get a bound on the acceleration of electron moving in $\theta$ coordinates to be $a_{\theta}\leq 10^{64}s^{-2}$. 
 
In order to calculate $a_{\theta}$, we set up a geodesic equation in DFR space-time, using star product formalism and derive its Newtonian limit. For solving Einstein's equation for linearised gravity, we derived the expression for Newtonian potential in the DFR space-time. This Newtonian potential is found to depend on the radial distance in $1+3+d_{\theta}$ dimensional DFR space-time as $\frac{1}{r^{1+d_{\theta}}}$. The Similar radial dependence of the Newtonian potential has also been observed in the study of gravity in extra dimensions \cite{extra}. Using this expression of Newtonian potential, we calculate $a_{\theta}$. This has then been used to put a lower bound on the radial distance between two gravitating bodies in DFR space-time. It is shown that this bound vanishes in the commutative limit as expected.

The MA expression in the commutative space-time has also been derived using Heisenberg's uncertainty principle \cite{caian-1984}. Adapting this approach and using the MA obtained in Eq.(\ref{timlikf}) for DFR space-time we find (see Appendix.A) that the MA leads to the modification of the uncertainty relation between the Minkowski coordinate and its conjugate in DFR space-time as $$[x_i,p_j]=i\hslash\Big(1-\frac{\lambda^2a_{\theta}^2\hslash^2}{4m^2c^6}\Big)\delta_{ij}.$$ This modification term depends on the minimal length scale $\lambda$, acceleration along the $\theta$ coordinate $a_{\theta}$, and rest mass of the particle $m$. Similar modifications were obtained in $\kappa$-deformed space-time \cite{kmax1} also. The generalised uncertainty principle is one of the common characteristics of microscopic models of gravity, and some of their implications have been worked out \cite{gup}. It will be quite interesting to see the implications of the above obtained modified commutation relation.

\section{Appendix A}
\renewcommand{\thesection}{A} 
In this appendix, we derive a length scale dependent commutation relation between commutative coordinate and its conjugate using the expression for the maximal acceleration obtained in the DFR space-time.

The uncertainty relation between energy and an arbitrary function of time $f(t)$ can be expressed as\cite{caian-1984}
\begin{equation}\label{0a}
  \Delta E\Delta f(t) \geq \frac{\hslash}{2}\frac{df(t)}{dt} 
\end{equation}  
Using the above relation we write down uncertainty relation between energy and velocity and that between energy and position as,   
\begin{equation}\label{a}
  \Delta E\Delta v(t) \geq \frac{\hslash}{2}\frac{dv}{dt} 
\end{equation}
and
\begin{equation}\label{v}
 \Delta E\Delta x(t) \geq \frac{\hslash}{2}\frac{dx}{dt}, 
\end{equation}
respectively. Multiplying the above uncertainty relations, given in Eq.(\ref{a}) and Eq.(\ref{v}), we get
\begin{equation}
 \big(\Delta E\big)^2\Delta x\frac{\Delta v}{v}\geq \frac{\hslash^2}{4}\mathcal{A}
\end{equation}
where we have used the definitions $v=\frac{dx}{dt}$ and $\mathcal{A}=\frac{dv}{dt}$. Now we express $\Delta E$ as $\Delta E=v\Delta p$, and we get
\begin{equation}
 \big(\Delta p\big)^2\Delta x\big({\Delta v}\big)v\geq \frac{\hslash^2}{4}\mathcal{A}.
\end{equation}
From the special theory of relativity we infer that the uncertainty in velocity of the particle cannot exceed the velocity of light, i.e, $(\Delta v)^2=<v^2>-<v>^2\leq v_{max}^2\leq c^2$, so we take $(\Delta v)v\leq c^2$. After using this relation in the above equation we find that 
\begin{equation}\label{un}
 \big(\Delta p\Delta x\big)^2c^2\geq \frac{\hslash^2}{4}\mathcal{A}\Delta x
\end{equation}
Re-writing $\mathcal{A}$ using $\mathcal{A}_{max}$ expression and setting $\Delta x$ in the RHS with Compton wavelength, $\lambda_c=\frac{\hslash}{mc}$  we get 
\begin{equation}\label{ku}
 \big(\Delta x\Delta p)^2 c^2\geq \frac{\hslash^2}{4}\frac{mc^3}{\hslash}\Big(1-\frac{\lambda^2a_{\theta}^2\hslash^2}{2m^2c^6}\Big)\frac{\hslash}{mc}.
\end{equation}
Thus we find that the commutation relation between the commutative coordinate and its conjugate momenta gets modified as
\begin{equation}\label{A2}
 [x_i,p_j]=i\hslash\Big(1-\frac{\lambda^2a_{\theta}^2\hslash^2}{4m^2c^6}\Big)\delta_{ij}.
\end{equation}
This shows that the existence of maximal acceleration implies a minimal length scale dependent modification to the commutation relation between coordinate and its conjugate. In \cite{kempf} a length scale dependent commutation relation known as GUP, between coordinate and its conjugate given by
\be
 [x_i,p_j]=i\hslash\Big(1+\frac{\beta l^2_{Pl}p^2}{\hslash^2}\Big)\delta_{ij}, \label{A3}
\ee
has been analysed and corresponding uncertainty relation between $x$ and $p$ was obtained. Comparing Eq.(\ref{A2}) with Eq.(\ref{A3}) and setting $\lambda=l_{Pl}$, we find the generalised uncertainty principle parameter $\beta$ in terms of the $a_{\theta}$ as $\beta=-\frac{a_{\theta}^2\hslash^4}{4m^2c^6p^2}$.  

\section*{Acknowledgements}
S.K.P.  thanks UGC, India for the support through JRF scheme (id.191620059604). V.R. thanks Government of India, for support through DST-INSPIRE/IF170622.

\bigskip



\end{document}